# Moderating Embodied Cyber Threats Using Generative AI

KEYAN GUO, University at Buffalo, USA

GUO FREEMAN, Clemson University, USA

HONGXIN HU, University at Buffalo, USA

## 1 INTRODUCTION

The advancement in computing and hardware, like spatial computing and VR headsets (e.g., Apple's Vision Pro) [1], has boosted the popularity of social VR platforms (VRchat, Rec Room, Meta Horizon Worlds) [2, 3, 4]. Unlike traditional digital interactions, social VR allows for more immersive experiences, with avatars that mimic users' real-time movements and enable physical-like interactions. However, the immersive nature of social VR may introduce intensified and more physicalized cyber threats–we define as *"embodied cyber threats"*, including trash-talking, virtual "groping", and such virtual harassment and assault. These new cyber threats are more realistic and invasive due to direct, virtual interactions, underscoring the urgent need for comprehensive understanding and practical strategies to enhance safety and security in virtual environments.

Generative AI (GenAI), such as ChatGPT [5], is able to understand, predict, and generate human-like responses, and has been widely used by researchers [6]. Building on our previous research, presented in Section 2, we highlight the potential of generative AI in addressing the complex challenge of embodied cyber threats within social VR environments. Three critical aspects of GenAI contribute to its effectiveness in combating embodied cyber threats in VR environments: First, GenAI's training involves extensive datasets covering a vast range of human interactions and behaviors, enabling it to understand and predict various user actions accurately. Secondly, GenAI's problem-solving is enhanced by chain-of-thought (CoT) reasoning, allowing it to break down and address complex issues in a human-like manner. Lastly, GenAI's ability to learn from human feedback ensures continuous adaptation and improvement, making it increasingly sensitive to the subtleties of social interactions and community norms in virtual spaces. Utilizing these strengths, GenAI can be harnessed to develop automated moderation tools to detect and mitigate unsafe behaviors in real-time, enhancing user safety and engagement in virtual environments. However, despite GenAI's potential in improving online safety, it may also pose risks, including the creation of new, more complex cyber threats. As these systems evolve, there's a possibility of misuse, leading to novel forms of virtual harassment or abuse. Specifically, sophisticated GenAI could craft highly realistic avatars or environments, potentially deceiving users and introducing harder-to-detect threats.

Building on these considerations, our interest lies in rigorously evaluating how GenAI can effectively moderate embodied cyber threats while identifying the dual-edged nature of such technology. Precisely, we aim to explore the opportunities GenAI presents in detecting and moderating embodied cyber threats in virtual environments, alongside the risks it may introduce, potentially exacerbating online harms rather than diminishing them. In such a case, we can investigate the balance between using GenAI for moderating harmful interactions and understanding how it could inadvertently amplify such threats. By identifying both the opportunities and challenges of employing GenAI, we intend to pave the way for developing robust, effective strategies that ensure GenAI aids in moderating online harms without spiraling out of control, maintaining a safe and inclusive environment in social VR.

## 2 PREVIOUS WORK

(1) *"Moderating Illicit Online Image Promotion for Unsafe User Generated Content Games Using Large Vision-Language Models."* Accept Conditional on Major Revision by **USENIX Security 2024**.
Online user-generated content games (UGCGs) like Roblox are popular among younger audiences but pose risks of exposure to explicit content. In this work, we addressed the challenge of identifying illicit UGCG promotion, a task complicated by the unique nature of UGCG images. We introduced UGCG-Guard, a GenAI-based system, specifically large vision-language models (VLMs), for efficient detection. UGCG-Guard employs conditional prompting for domain adaptation and CoT reasoning for contextual understanding and has achieved remarkable success in real-world testing.

(2) *"Moderating New Waves of Online Hate with Chain-of-Thought Reasoning in Large Language Models."* In **2024 IEEE Symposium on Security and Privacy (SP)**. [7]
In this work, we addressed the escalating issue of online hate, which is rapidly evolving and impacting Internet users worldwide. We proposed a novel framework, HateGuard, designed to mitigate this threat effectively. Utilizing delicate CoT prompting and GenAI, specifically large language models (LLMs), HateGuard dynamically updates detection prompts for new derogatory terms, addressing evolving online hate trends. Our validation involved a dataset from significant events, such as the 2022 Russian invasion of Ukraine, the 2021 US Capitol insurrection, and the COVID-19 pandemic, demonstrating HateGuard's superior performance in real-world scenarios.

(3) *"Towards Understanding and Detecting Cyberbullying in Real-world Images."* In **2021 Network and Distributed System Security Symposium (NDSS)**. [8]
This study explored cyberbullying through visual media, a less-studied aspect compared to textual content. We collected a large-scale dataset of real-world cyberbullying images and recognized five unique contextual factors after comprehensive analysis. Our findings showed distinct differences from traditional cyberbullying content, informing the development of effective classifier models. Our best model achieved a 93.36% detection accuracy, advancing our understanding and mitigating visual cyberbullying.

## 3 CONTRIBUTION TO THE WORKSHOP

In this workshop, we aim to explore the nuanced role of GenAI in moderating online harms within social VR environments. Our focus is on exploring GenAI's potential to enhance safety while addressing the challenges it poses in creating embodied cyber threats. Leveraging our research experience, we seek to contribute insights into developing effective approaches that maintain a balanced, safe, and inclusive digital space with GenAI. Our discussion will pivot around ethical deployment, best practices, and innovative approaches to ensure GenAI serves as a beneficial tool in the evolving landscape of social VR.